\documentclass[aps,prl,showpacs,twocolumn,lineno,groupedaddress]{revtex4}  
\usepackage{graphicx}  
\usepackage{dcolumn}   
\usepackage{bm}        
\usepackage{amssymb}   

\def\MET{{\mbox{$E\kern-0.57em\raise0.19ex\hbox{/}_{T}$}}}
\def\DZero{D\O\ }

\begin{document}


\title{Search for Supersymmetry with Gauge-Mediated Breaking in Diphoton Events
at \DZero }
%
\author{                                                                      
V.M.~Abazov,$^{32}$                                                           
B.~Abbott,$^{69}$                                                             
M.~Abolins,$^{60}$                                                            
B.S.~Acharya,$^{26}$                                                          
D.L.~Adams,$^{67}$                                                            
M.~Adams,$^{47}$                                                              
T.~Adams,$^{45}$                                                              
M.~Agelou,$^{16}$                                                             
J.-L.~Agram,$^{17}$                                                           
S.N.~Ahmed,$^{31}$
% ...
\\                                                                            
\vskip 0.30cm                                                                 
\centerline{(D\O\ Collaboration)}                                             
\vskip 0.30cm                                                                 
}                                                                             
\address{                                                                     
\centerline{$^{1}$Universidad de Buenos Aires, Buenos Aires, Argentina}       
\centerline{$^{2}$LAFEX, Centro Brasileiro de Pesquisas F{\'\i}sicas,         
                  Rio de Janeiro, Brazil}                                     
\centerline{$^{3}$Universidade do Estado do Rio de Janeiro,                   
                  Rio de Janeiro, Brazil}                                     
\centerline{$^{4}$Instituto de F\'{\i}sica Te\'orica, Universidade            
                  Estadual Paulista, S\~ao Paulo, Brazil}                     
\centerline{$^{5}$University of Alberta and Simon Fraser University,          
                  Canada}                                                     
\centerline{$^{6}$Institute of High Energy Physics, Beijing,                  
                  People's Republic of China}                                 
\centerline{$^{7}$Universidad de los Andes, Bogot\'{a}, Colombia}             
\centerline{$^{8}$Charles University, Center for Particle Physics,            
                  Prague, Czech Republic}                                     
\centerline{$^{9}$Czech Technical University, Prague, Czech Republic}         
\centerline{$^{10}$Institute of Physics, Academy of Sciences, Center          
                  for Particle Physics, Prague, Czech Republic}               
\centerline{$^{11}$Universidad San Francisco de Quito, Quito, Ecuador}        
\centerline{$^{12}$Laboratoire de Physique Subatomique et de Cosmologie,      
                  IN2P3-CNRS, Universite de Grenoble 1, Grenoble, France}     
\centerline{$^{13}$CPPM, IN2P3-CNRS, Universit\'e de la M\'editerran\'ee,     
                  Marseille, France}                                          
\centerline{$^{14}$Laboratoire de l'Acc\'el\'erateur Lin\'eaire,              
                  IN2P3-CNRS, Orsay, France}                                  
\centerline{$^{15}$LPNHE, Universit\'es Paris VI and VII, IN2P3-CNRS,         
                  Paris, France}                                              
\centerline{$^{16}$DAPNIA/Service de Physique des Particules, CEA, Saclay,    
                  France}                                                     
\centerline{$^{17}$IReS, IN2P3-CNRS, Univ. Louis Pasteur Strasbourg,          
                   and Univ. de Haute Alsace, France}                         
\centerline{$^{18}$Institut de Physique Nucl\'eaire de Lyon, IN2P3-CNRS,      
                   Universit\'e Claude Bernard, Villeurbanne, France}         
\centerline{$^{19}$RWTH Aachen, III. Physikalisches Institut A,               
                   Aachen, Germany}                                           
\centerline{$^{20}$Universit{\"a}t Bonn, Physikalisches Institut,             
                  Bonn, Germany}                                              
\centerline{$^{21}$Universit{\"a}t Freiburg, Physikalisches Institut,         
                  Freiburg, Germany}                                          
\centerline{$^{22}$Universit{\"a}t Mainz, Institut f{\"u}r Physik,            
                  Mainz, Germany}                                             
\centerline{$^{23}$Ludwig-Maximilians-Universit{\"a}t M{\"u}nchen,            
                   M{\"u}nchen, Germany}                                      
\centerline{$^{24}$Fachbereich Physik, University of Wuppertal,               
                   Wuppertal, Germany}                                        
\centerline{$^{25}$Panjab University, Chandigarh, India}                      
\centerline{$^{26}$Tata Institute of Fundamental Research, Mumbai, India}     
\centerline{$^{27}$University College Dublin, Dublin, Ireland}                
\centerline{$^{28}$Korea Detector Laboratory, Korea University,               
                   Seoul, Korea}                                              
\centerline{$^{29}$CINVESTAV, Mexico City, Mexico}                            
\centerline{$^{30}$FOM-Institute NIKHEF and University of                     
                  Amsterdam/NIKHEF, Amsterdam, The Netherlands}               
\centerline{$^{31}$University of Nijmegen/NIKHEF, Nijmegen, The               
                  Netherlands}                                                
\centerline{$^{32}$Joint Institute for Nuclear Research, Dubna, Russia}       
\centerline{$^{33}$Institute for Theoretical and Experimental Physics,        
                  Moscow, Russia}                                             
\centerline{$^{34}$Moscow State University, Moscow, Russia}                   
\centerline{$^{35}$Institute for High Energy Physics, Protvino, Russia}       
\centerline{$^{36}$Petersburg Nuclear Physics Institute,                      
                   St. Petersburg, Russia}                                    
\centerline{$^{37}$Lund University, Royal Institute of Technology,            
                   Stockholm University, and Uppsala University, Sweden}      
\centerline{$^{38}$Lancaster University, Lancaster, United Kingdom}           
\centerline{$^{39}$Imperial College, London, United Kingdom}                  
\centerline{$^{40}$University of Manchester, Manchester, United Kingdom}      
\centerline{$^{41}$University of Arizona, Tucson, Arizona 85721}              
\centerline{$^{42}$Lawrence Berkeley National Laboratory and University of    
                  California, Berkeley, California 94720}                     
\centerline{$^{43}$California State University, Fresno, California 93740}     
\centerline{$^{44}$University of California, Riverside, California 92521}     
\centerline{$^{45}$Florida State University, Tallahassee, Florida 32306}      
\centerline{$^{46}$Fermi National Accelerator Laboratory, Batavia,            
                   Illinois 60510}                                            
\centerline{$^{47}$University of Illinois at Chicago, Chicago,                
                   Illinois 60607}                                            
\centerline{$^{48}$Northern Illinois University, DeKalb, Illinois 60115}      
\centerline{$^{49}$Northwestern University, Evanston, Illinois 60208}         
\centerline{$^{50}$Indiana University, Bloomington, Indiana 47405}            
\centerline{$^{51}$University of Notre Dame, Notre Dame, Indiana 46556}       
\centerline{$^{52}$Iowa State University, Ames, Iowa 50011}                   
\centerline{$^{53}$University of Kansas, Lawrence, Kansas 66045}              
\centerline{$^{54}$Kansas State University, Manhattan, Kansas 66506}          
\centerline{$^{55}$Louisiana Tech University, Ruston, Louisiana 71272}        
\centerline{$^{56}$University of Maryland, College Park, Maryland 20742}      
\centerline{$^{57}$Boston University, Boston, Massachusetts 02215}            
\centerline{$^{58}$Northeastern University, Boston, Massachusetts 02115}      
\centerline{$^{59}$University of Michigan, Ann Arbor, Michigan 48109}         
\centerline{$^{60}$Michigan State University, East Lansing, Michigan 48824}   
\centerline{$^{61}$University of Mississippi, University, Mississippi 38677}  
\centerline{$^{62}$University of Nebraska, Lincoln, Nebraska 68588}           
\centerline{$^{63}$Princeton University, Princeton, New Jersey 08544}         
\centerline{$^{64}$Columbia University, New York, New York 10027}             
\centerline{$^{65}$University of Rochester, Rochester, New York 14627}        
\centerline{$^{66}$State University of New York, Stony Brook,                 
                   New York 11794}                                            
\centerline{$^{67}$Brookhaven National Laboratory, Upton, New York 11973}     
\centerline{$^{68}$Langston University, Langston, Oklahoma 73050}             
\centerline{$^{69}$University of Oklahoma, Norman, Oklahoma 73019}            
\centerline{$^{70}$Brown University, Providence, Rhode Island 02912}          
\centerline{$^{71}$University of Texas, Arlington, Texas 76019}               
\centerline{$^{72}$Rice University, Houston, Texas 77005}                     
\centerline{$^{73}$University of Virginia, Charlottesville, Virginia 22901}   
\centerline{$^{74}$University of Washington, Seattle, Washington 98195}       
}                                                                             
\date{\today}

\begin{abstract}
We report the results of a search for supersymmetry (SUSY) with
gauge-mediated breaking in the missing transverse
energy distribution of inclusive diphoton events
using  263 pb$^{-1}$ of data collected by the
D\O\ experiment at the Fermilab Tevatron Collider in 2002--2004. No excess
is observed above the background expected from 
standard model processes, and lower limits on the masses of the lightest neutralino and
chargino of about 108 and 195 GeV,  respectively,  are set at the
95\% confidence level. These are the most stringent limits to date for
models with gauge-mediated SUSY breaking
with a short-lived neutralino as the next-lightest SUSY particle.
\end{abstract}
\maketitle

Models involving gauge-mediated supersymmetry breaking
(GMSB), originally proposed in Ref. \cite{gmsb_original} 
have attracted much attention \cite{gmsbsusy}.
In GMSB models supersymmetry breaking is achieved by introduction
of new chiral supermultiplets, called messengers,
which couple to the ultimate source of supersymmetry breaking, and
also to the SUSY particles.
The phenomenology of these models is rich
and strikingly different from that of gravity-mediated SUSY models.

For GMSB models,  the gravitino (with a mass less than $\sim ~ $keV) is the
lightest SUSY particle (LSP),  and the phenomenology of these
models is therefore determined by the next-to-lightest SUSY particle
(NLSP),  which can be either a neutralino or a slepton. 
In the former case, which is considered in this paper,
the NLSP decays into a photon and an LSP, and the signal of interest,
assuming $R$-parity conservation \cite{rpar}, is a final
state with two photons and large missing transverse energy
($\MET$). 

The model we consider is a minimal GMSB
with a neutralino as the NLSP, referred to as Snowmass Slope SPS 8
\cite{modelline}. This model has only one dimensioned parameter
$\Lambda$ that determines the effective
scale of SUSY breaking. The minimal GMSB parameters correspond to a messenger
mass $M_m = 2\Lambda$, the number of messengers $N_5 = 1$,  the ratio of
the vacuum expectation values of the two Higgs fields $\tan \beta
= 15$,  and the sign of the Higgsino mass term $\mu > 0$.
The lifetime of the neutralino is
not fixed by this model line, and is assumed to be sufficiently short to
result in decays with prompt photons.
Current lower limits on the GMSB neutralino mass for somewhat similar model 
parameters are 65,  75 and 100 GeV,  from the 
CDF \cite{cdfrun1},  D\O\ \cite{d0run1} and CERN LEP
collaborations \cite{lepsusy},  respectively.

We search for SUSY production in $p\bar{p}$ collisions at
$\sqrt{s} = 1.96$ TeV at the Fermilab Tevatron Collider.
The D\O\ detector comprises a central tracking system in 
a 2 T superconducting solenoidal magnet, a liquid-argon/uranium 
calorimeter, and a muon spectrometer~\cite{run2det}. 
The tracking system consists of a silicon microstrip tracker and
a scintillating fiber tracker and provides coverage for charged particles
in the pseudorapidity range $|\eta| < 3$.
The calorimeters are finely segmented and consist of a central 
section (CC) covering 
$|\eta| \leq 1.1$, and two end calorimeters (EC) 
extending coverage to $|\eta|\approx 4$, all housed in separate 
cryostats~\cite{run1det}.
Scintillators installed between the CC and EC cryostats provide sampling 
of developing showers for $1.1<|\eta|<1.4$.
The electromagnetic (EM) section of the calorimeter has four longitudinal
layers and transverse segmentation of $0.1 \times 0.1$ in $\eta - \phi$
space (where $\phi$ is the azimuthal angle), except in the third layer, 
corresponding to EM shower maximum, where it is $0.05 \times 0.05$.
The data sample was collected
between April 2002 and March 2004, using inclusive single
electromagnetic (EM) and di-EM triggers. The integrated luminosity of
the sample is $263 \pm 17 ~ \rm pb^{-1}$.

Photons and electrons are identified in two steps: first, selection of
the EM clusters, and then their separation into photons or electrons. 
EM clusters are selected from calorimeter clusters  by requiring that (i)
at least 90\% of the energy be deposited in the EM section of the 
calorimeter, (ii) the calorimeter isolation variable ($I$) be 
less than 0.15, where 
$I = [E_{tot}(0.4)-E_{EM}(0.2)] / E_{EM}(0.2)$, where
$E_{tot}(0.4)$ is the total shower energy in a cone of radius
${\cal R} = \sqrt{(\Delta\eta)^2 + (\Delta\phi)^2} = 0.4$, and 
$E_{EM}(0.2)$ is the EM energy in a cone ${\cal R}=0.2$,
 (iii) the transverse and longitudinal shower profiles be
consistent with those expected for an EM shower, and (iv) 
the scalar sum of the $p_T$ of all tracks originating from the primary 
vertex in an annulus of $ 0.05 < {\cal R} < 0.4$ around the cluster be 
less than 2 GeV. The cluster is then defined as an electron if there
is a reconstructed track pointing to it and a photon otherwise. 
Jets are reconstructed using the iterative, midpoint cone algorithm \cite{run2cone} with 
a cone size of 0.5. $\MET$ is determined 
from the energy deposited in the calorimeter for $|\eta| < 4$ and is
corrected for jet and EM energy scales.

We select $\gamma\gamma$ candidates by requiring
events to have two photons each with $E_T > 20~$GeV and
pseudorapidity $|\eta|<1.1$. To suppress events with mismeasured
$\MET$, we apply the following requirements.
We reject any event when the difference in azimuth ($\Delta \phi$) between the highest
$E_T$ jet (if jets are present) and the direction of the $\MET$ is more than 2.5
radians, or if the $\Delta \phi$ between the direction of 
the $\MET$ and either photon is less than 0.5 radians. 
These selections yield 1,909 events ($\gamma\gamma$ sample), 
out of which 1,800 have $\MET <
15~$GeV and two have $\MET > 40~$GeV. The two events
constitute the $\gamma\gamma\MET$ sample.

\begin{figure}
\includegraphics[scale=0.4]{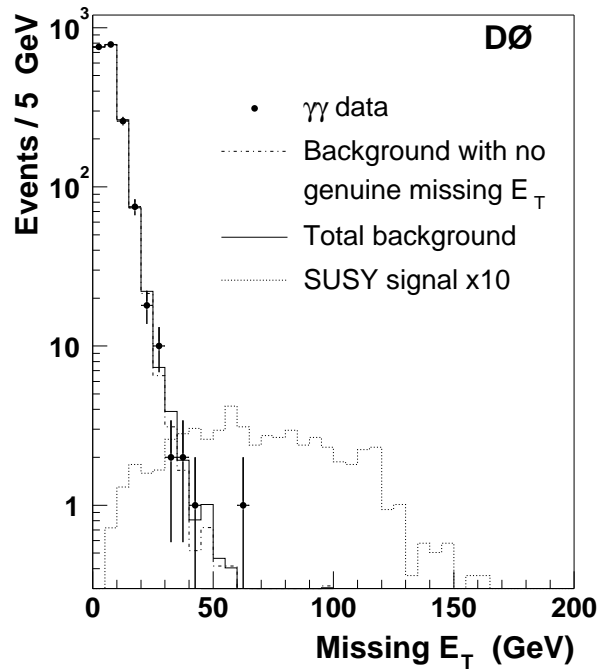}
\caption{\label{fig_gg} The $\MET$ distribution for the diphoton 
and background samples. Also shown is the expected distribution for
the GMSB point with $\Lambda = 80~ $TeV, multiplied by a factor of ten.}
\end{figure}

The main backgrounds arise from standard model processes with
misidentified photons and/or mismeasured $\MET$. 
The background from processes with no inherent $\MET$
(multijet events, direct photon production, $Z \rightarrow ee$, etc.)
is estimated using events with two EM clusters that satisfy
photon-identification criteria (i) and (ii), but fail the shower-shape requirement
(iii). These events, called the QCD sample, must pass
the same trigger and other selections that define the $\gamma\gamma$ sample.
They have characteristics similar to the 
background in the $\gamma\gamma$ sample 
and in particular are expected to have similar $\MET$ resolution. 
This assumption was checked by varying the selection criteria
and comparing the $\MET$ distribution in the QCD sample 
to that in $Z \rightarrow ee$ events. 
The QCD sample comprises 18,437 events, with 17,379 events having 
$\MET < 15 ~$GeV, and 27 events with $\MET > 40 ~$GeV.
We estimate the background in the $\gamma\gamma\MET$ sample 
resulting from  mismeasurement of $\MET$ by 
normalizing the number of QCD events to that of the $\gamma\gamma$ sample  
for $\MET < 15~$GeV.
This yields $2.8 \pm 0.5$ events with $\MET > 40 ~$GeV,
with uncertainty dominated by the statistics of the QCD sample.

The other sources of background correspond to events with genuine $\MET$ in which
an electron is misidentified as a photon, for example
 from $W + $\mbox{'$\gamma$'} events
(where \mbox{'$\gamma$'} denotes both true photons and jets misidentified as photons), 
and from $Z \rightarrow \tau^+\tau^- \rightarrow e^+e^- + X$ and $t\bar{t}
\rightarrow e^+e^- + ~$jets production. We estimate this contribution
using the $e\gamma$ sample which has the same trigger, kinematic,
and EM identification requirements as the $\gamma\gamma$
sample. This sample contains 889 events, 782 events with $\MET < 15 ~$GeV and 15 
events with $\MET > 40 ~$GeV. To estimate the contribution of such events
to the $\gamma\gamma\MET$ sample, we first subtract the QCD
background component of the $e\gamma$ sample. This is done by
normalizing the QCD sample to the $e\gamma$ sample for $\MET <
15~$GeV. Then, using the probability for an electron to be misidentified as
a photon (measured using $Z \rightarrow ee$ events to be 
$0.064 \pm 0.004$), we estimate this background to be $0.9 \pm 0.2$
events with statistically dominated uncertainty.
Therefore the total expected background to the $\gamma\gamma\MET$ sample
is $3.7 \pm 0.6$ events. The $\MET$ distributions for
the $\gamma\gamma$ sample, background without genuine $\MET$, and
the total background 
are shown in Fig. \ref{fig_gg}, together with an expected 
distribution from the Snowmass Slope model with $\Lambda = 80~$TeV,
the latter multiplied by a factor of ten for clarity.

To estimate the expected signal, we generated 
Monte Carlo (MC) events for several points on the Snowmass Slope (see Table
\ref{table_gmsb}), covering the neutralino mass range from $72~$GeV,
somewhat below the existing limits \cite{d0run1, lepsusy}, to $116~$GeV.
We used {\sc isajet 7.58} \cite{isajet} to
 determine SUSY interaction eigenstate masses and couplings.
{\sc pythia 6.202} \cite{pythia} 
was used to generate the events after determining the sparticle masses,
branching fractions and leading order (LO) production cross sections using
the {\sc CTEQ5L} \cite{cteq} parton distribution functions (PDF).
MC events were processed through full detector simulation 
and reconstruction, and processed with the analysis program used for the data.

The dominant contributions to the cross section 
are from production of lightest charginos ($\tilde{\chi}^+_1\tilde{\chi}^-_1$)
and chargino-second neutralino pairs ($\tilde{\chi}^0_2\tilde{\chi}^{\pm}_1$).
The total cross section in Table \ref{table_gmsb} is calculated 
to leading order in {\sc pythia}  for GMSB SUSY
production. The ``$K$-factor'' used to 
account for higher-order corrections is applied to estimate the 
next-to-leading-order (NLO) cross section. The values of the 
$K$-factor in the table are taken from Ref. \cite{kfact}.
The sources of error on signal efficiency include uncertainty on photon
identification (4\% per photon), MC statistics (5\%), and choice of
PDF (5\%). 

\begin{table*}
\caption{ \label{table_gmsb}
Points on the Snowmass Slope: their cross sections,
efficiencies and cross-section limits.}
\begin{ruledtabular}
\begin{tabular}{c|cc|cc|c|c}
$\Lambda,$ TeV &$m_{\tilde{\chi}_1^0},$GeV & $m_{\tilde{\chi}_1^+},$GeV & $\sigma_{TOT}^{LO}$, pb & $K$-factor & Efficiency & 95\% C.L. Limit, pb\\ 
\hline
 55 & 71.8 & 126.3 & 0.735 & 1.236 & $0.092\pm 0.009$ & 0.184 \\ 
 60 & 79.1 & 140.2 & 0.468 & 1.227 & $0.100\pm 0.009$ & 0.170 \\ 
 65 & 86.4 & 154.3 & 0.301 & 1.217 & $0.111\pm 0.011$ & 0.153 \\ 
 70 & 93.7 & 168.2 & 0.204 & 1.207 & $0.124\pm 0.012$ & 0.137 \\ 
 75 & 101.0 & 182.3 & 0.138 & 1.197 & $0.137\pm 0.013$ & 0.124 \\ 
 80 & 108.2 & 196.0 & 0.094 & 1.187 & $0.149\pm 0.014$ & 0.114 \\ 
 85 & 115.5 & 209.9 & 0.066 & 1.177 & $0.154\pm 0.015$ & 0.110 \\ 
\end{tabular}
\end{ruledtabular}
\end{table*}

\begin{table}
\caption{ \label{table_limit}
Limits on the Snowmass Slope and two other GMSB models. }
\begin{ruledtabular}
\begin{tabular}{cccc|cccc}
\multicolumn{4}{c}{Fixed parameters} & \multicolumn{4}{c}{95\% CL lower limits} \\
 $M_m / \Lambda$ & $\tan \beta$ & $N_5$ &
sign($\mu$) & $\Lambda $ & $m_{\tilde{\chi}_1^0}$ &
$m_{\tilde{\chi}_1^+}$ &
$m_{\tilde{\chi}_2^0}$ \\ 
\hline
 2  & 15 & 1 &  $+$ & 79.6 & 107.7 & 194.9   & 195.9 \\ 
 2  &  5 & 1 &  $+$ & 79.5 & 106.0 & 191.6   & 193.3 \\ 
 10 &  5 & 2 &  $+$ & 44.0 & 111.4 & 196.0   & 198.7 \\ 
\end{tabular}
\end{ruledtabular}
\end{table}

\begin{figure}
\includegraphics[scale=0.4]{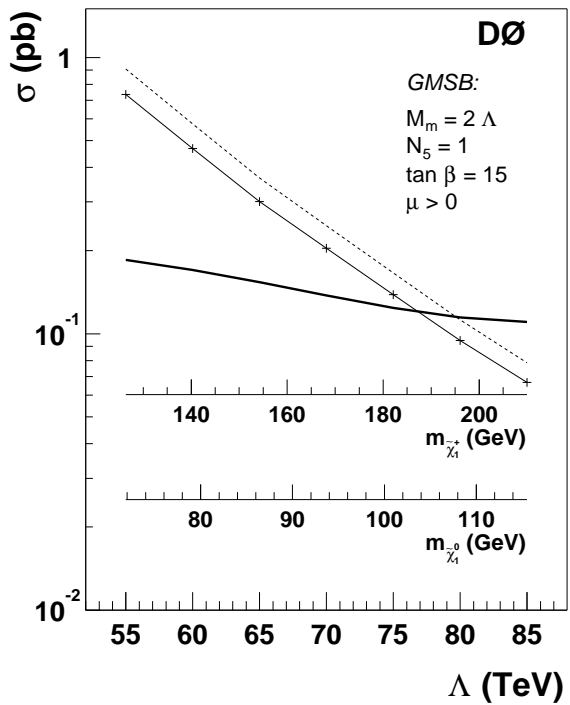}
\caption{\label{fig_limit} Predicted cross sections for the Snowmass Slope model vs $\Lambda$ in leading order (thin solid line with crosses), multiplied by the $K$-factor (thin dashed line),
and the 95\% C.L. limits (solid line).}
\end{figure}

Since the observed number of events is in good agreement with that
expected from the standard model, we conclude that there is no evidence
for GMSB SUSY in our data. To calculate the upper
limit on the production cross section for each sampled point on the Snowmass
Slope, 
we use a Bayesian approach \cite{bayesian} 
with a flat prior for the signal cross section. The
calculation takes into account uncertainties on the expected number of background
events, efficiency, and luminosity. 
The selection $\MET > 40~$ GeV for the signal sample leads to the best
expected limit, given the predicted background and expected signal distributions.
Our limits are shown in Table 
\ref{table_gmsb}, and plotted in Fig. \ref{fig_limit}, together with
the expected signal cross section.
The upper limit on the cross section is
below the expected value for $\Lambda <
79.6~ $TeV, corresponding to lower limits on gaugino masses of
$m_{\tilde{\chi}^+_1} > 194.9~$GeV and $m_{\tilde{\chi}^0_1} >
107.7~$GeV. The expected limit, given the predicted number of 
background events, is $\Lambda > 74.5~ $TeV.
We find that the gaugino mass limits depend only slightly on 
the parameters of the minimal GMSB. We have considered  models with
values of $\tan \beta$ and $N_5$ different from the Snowmass
Slope, and arrive at very similar results as detailed by Table \ref{table_limit}.

To summarize, we searched for inclusive high-$E_T$ diphoton events with large
missing transverse energy. Such events are predicted in 
supersymmetric models with low-scale gauge-mediated supersymmetry
breaking. We find no excess of such events, and interpret the result
as a lower limit on gaugino masses. For a representative point in
the parameter space, we determine that at a 95\% confidence level, the
masses of the lightest chargino and neutralino are larger than
195 and 108 GeV, respectively. These are the most restrictive
limits to date for the Snowmass Slope model.

We thank S. Martin for valuable discussions and 
S. Mrenna for his help with the event generators.
%
We thank the staffs at Fermilab and collaborating institutions, 
and acknowledge support from the 
Department of Energy and National Science Foundation (USA),  
Commissariat  \` a l'Energie Atomique and 
CNRS/Institut National de Physique Nucl\'eaire et 
de Physique des Particules (France), 
Ministry of Education and Science, Agency for Atomic 
   Energy and RF President Grants Program (Russia),
CAPES, CNPq, FAPERJ, FAPESP and FUNDUNESP (Brazil),
Departments of Atomic Energy and Science and Technology (India),
Colciencias (Colombia),
CONACyT (Mexico),
KRF (Korea),
CONICET and UBACyT (Argentina),
The Foundation for Fundamental Research on Matter (The Netherlands),
PPARC (United Kingdom),
Ministry of Education (Czech Republic),
Natural Sciences and Engineering Research Council and 
WestGrid Project (Canada),
BMBF and DFG (Germany),
A.P.~Sloan Foundation,
Civilian Research and Development Foundation,
Research Corporation,
Texas Advanced Research Program,
and the Alexander von Humboldt Foundation.
%

\end{document}